\documentclass[twocolumn,apl,showpacs,showkeys,preprintnumbers,amsmath,amssymb]{revtex4}
\usepackage{graphicx}% Include figure files
\usepackage{dcolumn}% Align table columns on decimal point
\usepackage{bm}% bold math

\begin{document}
\preprint{Feng et. al., submitted to Appl. Phys. Lett.}

\title{A novel route to lower the dielectric loss in CaCu$_{3}$Ti$_{4}$O$_{12}$ ceramics}

\author{Lixin Feng,$^{1}$ Xiaoming Tang,$^{2}$ Yueyue Yan,$^{1}$ Xuezhi Chen,$^{1}$ Zhengkuan Jiao,$^{1}$ Guanghan Cao$^{1}$\footnote[1]{To whom correspondence should be addressed
(ghcao@zju.edu.cn)}}

\affiliation{$^{1}$Department of Physics, Zhejiang University,
Hangzhou 310027, People's Republic of China}

\affiliation{$^{2}$Test and Analysis Center, Zhejiang University,
Hangzhou 310027, People's Republic of China}

\begin{abstract}
Based upon a widely-accepted internal-barrier-layer-capacitance
(IBLC) model, the dielectric loss in the giant-dielectric-constant
CaCu$_{3}$Ti$_{4}$O$_{12}$ (CCTO) material is expected to decrease
as the conductivity of the CCTO grains/subgrains increases. In
this letter, we report that this idea was successfully realized
through the La-for-Ca substitution in the CCTO ceramics. The
impedance spectroscopy analysis gives the evidence.
\end{abstract}



\maketitle

The perovskite-like material CaCu$_{3}$Ti$_{4}$O$_{12}$ (CCTO) has
attracted much attention during the last few years due to its
extraordinary dielectric properties. It exhibits very high value
of low-frequency dielectric constant ($\varepsilon_r \sim$ 10,000)
over a wide temperature range (100 - 500 K), while the crystal
structure remains cubic and centric above 35
K.\cite{Subramanian,Ramirez} Furthermore, the dielectric constant
is almost temperature independent from 100 K to 400 K, which is
desirable for many micro-electric applications. However, the
dissipation factor of the material is relatively too high. For the
ceramic\cite{Subramanian} and thin-film\cite{Si} samples at room
temperature, the typical value of tan$\delta$ is about 0.1 at 1
kHz.

As we know, the dielectric loss is closely related with the
mechanism of the dielectric response. Subramanian \emph{et
al}.\cite{Subramanian} initially proposed that the dielectric
constant of CCTO might be enhanced by a barrier layer mechanism,
assuming that there was a naturally-formed inhomogeneous
microstructure such as twin boundaries. Later, Sinclair \emph{et
al}.\cite{Sinclair} confirmed this idea by the impedance
spectroscopy measurement which demonstrated that the CCTO ceramics
were electrically heterogeneous and consisted of semiconducting
grains with insulating grain boundaries. Very recently, by using
microcontact current-voltage measurements Chung \emph{et
al}.\cite{Chung} found a strong nonlinear current-voltage
behaviour in the undoped CCTO ceramics, suggesting an
electrostatic barrier at the grain boundaries. Nevertheless, the
insulating barrier should not be only the grain boundaries but
also some kinds of extended planar defects inside the grains,
because exceptionally high $\varepsilon_r$ was observed in CCTO
crystals.\cite{Homes-a,Li} On the whole, the IBLC model with the
Maxwell-Wagner relaxation is widely
accepted\cite{Lunkenheimer,Cohen,Homes-b,Liu} as the primary
mechanism for the appearance of giant dielectric constant in CCTO
and its related compounds.

Based on the IBLC model, the dielectric loss mainly originates
from the conductivity of the CCTO conducting crystalline
grains/subgrains as well as that of the insulating barriers. The
conductance of the barriers leads to the leakage loss. On the
other hand, the alternate current within the conducting region
also produces a dissipation as the Joule heat. Unlike the leakage
loss that can be suppressed by increasing the resistance of the
barriers, \emph{the dissipation loss from the conducting regions
is expected to decrease as the conductance of the CCTO
grains/subgrains increases}. Therefore, we performed a partial
substitution of divalent Ca$^{2+}$ by the trivalent La$^{3+}$ in
order to increase the conductivity of the grains/subgrains. Our
result demonstrates that the dielectric loss of the CCTO ceramics
was remarkably lowered by the La substitution, while the giant
dielectric property still remains.

Ceramic samples of Ca$_{1-x}$La$_{x}$Cu$_{3}$Ti$_{4}$O$_{12}$ ($x$
= 0, 0.1, 0.2, 0.3 and 0.4) were prepared by a conventional
solid-state reaction. The starting materials were high purity
(99.99\%) CaCO$_3$, CuO, La$_2$O$_3$ and TiO$_2$. They were
weighed according to the stoichiometric ratios and mixed
thoroughly in an agate mortar. The mixture was calcined at 1223 K
for 12 h in air. Then the calcined powder was reground and pressed
into disks of 10 mm in diameter and 2 mm in thickness. The disks
were sintered in air at about 1333 K for 24 h and furnace-cooled
to room temperature. The sintered temperature varied a little for
different samples to obtain similar density and grain size. The
scanning electron microscopy images (not shown here) for the
samples' cross sections indicated that the samples had the similar
morphology and grain size (about 5 $\mu$m). This observation
ensures the comparability for the sample's dielectric properties.

The dielectric property was measured using an Agilent 4284A
precision LCR meter. The capacitance $C_p$ and the dissipation
factor $D$ (i.e., tan$\delta$) of the parallel-plate capacitor
made of the disk sample were directly measured at an ac voltage of
1 V in the frequency range of 20 - 10$^6$ Hz and the temperature
range of 80 - 300 K. The $\varepsilon_r$ value was easily obtained
from $C_p$ and the sample's size (the thickness and the area of
the electrodes). The complex impedance was obtained by the
equation $Z^{*}=1/(i\omega C^{*})$, where $C^{*}=C_{p}-iDC_{p}$
and $\omega=2\pi f$.

Figure 1 shows the powder x-ray diffraction (XRD) results for the
samples of Ca$_{1-x}$La$_{x}$Cu$_{3}$Ti$_{4}$O$_{12}$. Except for
$x$ = 0.4, all the XRD peaks of each sample can be well indexed
using a body-centered cubic lattice with $a \sim $ 0.74 nm. The
lattice constant, obtained by a least-squares fit with the
consideration of zero-shift, increases monotonically with
increasing $x$ (see the inset), confirming that La does enter the
lattice. When $x\geq$ 0.3, the lattice constant tends to saturate,
suggesting the La-for-Ca solubility limit at $x \sim $ 0.3. This
result is in agreement with the observation that the sample of $x$
= 0.4 contains small amount of secondary phase marked by an
asterisk. The increase of the lattice constant is probably due to
the fact that La$^{3+}$ has larger ionic radius than Ca$^{2+}$
does. Besides, the possible formation of larger Ti$^{3+}$
(compared with Ti$^{4+}$) due to the substitution might also play
a role for expanding the lattice. One notes that the x-ray
photoemission spectroscopy indicated the existence of Ti$^{3+}$ in
CCTO.\cite{Zhang}

\begin{figure}[tbp]
\includegraphics[width=7.5cm]{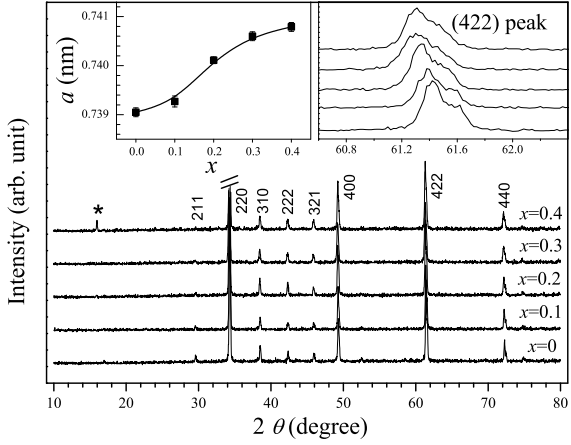}
\caption{Powder XRD results of
Ca$_{1-x}$La$_{x}$Cu$_{3}$Ti$_{4}$O$_{12}$ ceramic samples. The
top right inset is an expanded view of the XRD patterns around
2$\theta$ = 61.4$^{\circ}$, showing the shift of the (422) peaks.
The top left one shows the lattice constant as a function of $x$.}
\end{figure}

Figure 2 shows the temperature and frequency dependences of the
dielectric constant and the loss tangent for the
Ca$_{1-x}$La$_{x}$Cu$_{3}$Ti$_{4}$O$_{12}$ monophasic ceramics. As
can be seen, all the samples show giant dielectric constant of
$\varepsilon_r \geq$ 3000 in broad temperature and frequency
ranges though the dielectric constant decreases to some extent
with the La substitution. In addition, the La-substituted samples
have the temperature-independent $\varepsilon_r$ down to lower
temperature and up to higher frequency. The striking effect of the
La substitution comes from the decrease of tan$\delta$. As can be
seen in figure 2(b), the tan$\delta$ values in the whole measured
temperature range are decreased by the La substitution. At the
lowly-substituting level up to $x$ = 0.1, this effect is rather
remarkable. While from $x$ = 0.1 to $x$ = 0.2, the magnitude of
the decrease becomes smaller. And for samples of $x$ = 0.2 and
0.3, the dielectric loss is nearly identical. Compared back with
the parent CCTO sample, the dissipation factor of the $x$ = 0.2
sample is decreased by 3 to 5 times. The frequency dependence of
tan$\delta$ shown in figure 2(d) indicates that the most prominent
decrease of tan$\delta$ occurs at the intermediate frequencies.
For example, the tan$\delta$ value at 1 kHz is decreased from
$\sim$ 0.1 to $\sim$ 0.015 as the La substitutes for Ca at $x$ =
0.2.

\begin{figure}[tbp]
\includegraphics[width=7.5cm]{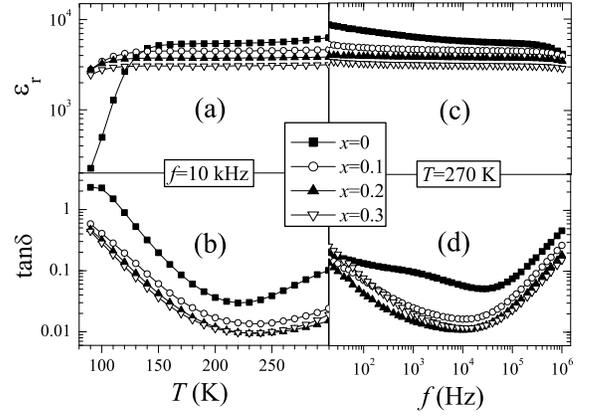}
\caption{La-for-Ca substitution effect on the dielectric
properties in the Ca$_{1-x}$La$_{x}$Cu$_{3}$Ti$_{4}$O$_{12}$
ceramics.}
\end{figure}

To understand the La substitution effect on the dielectric loss,
let us first derive the expression of tan$\delta$ under the IBLC
mechanism. Considered the case of single barrier, for simplicity,
the CCTO system can be modelled by an equivalent circuit
consisting of two parallel $RC$ elements connected in series. The
two parallel $RC$ units represent semiconducting regions
($R_{sc}$,$C_{sc}$) and insulating barriers ($R_{ins}$,$C_{ins}$),
respectively. One may easily write out the expression of the total
complex impedance as,
\begin{equation}
Z^{*}=\frac{R_{sc}}{1+i\omega
R_{sc}C_{sc}}+\frac{R_{ins}}{1+i\omega R_{ins}C_{ins}}.
\label{eq1}
\end{equation}
From the above equation, the complex admittance, known as the
reciprocal of $Z^{*}$, can be derived. Since $R_{ins} \gg R_{sc}$
and $C_{ins} \gg C_{sc}$ in the CCTO system,\cite{Sinclair} the
real part of the admittance (i.e., the conductance $G$) can be
simplified as,
\begin{equation}
G=\frac{1+\omega^{2}R_{sc}R_{ins}C_{ins}^{2}}{R_{ins}+\omega^{2}
R_{sc}^{2}R_{ins}C_{ins}^{2}}. \label{eq2}
\end{equation}
Noted that $\omega R_{sc}C_{ins} \ll 1$ is basically satisfied in
our measured frequency range, and $C_{ins} \approx C_{p}$, the
loss tangent (tan$\delta=G/(\omega C_{p})$) is then approximately
to be,
\begin{equation}
\mathrm{tan}\delta=\frac{1}{\omega R_{ins}C_{p}}+\omega
R_{sc}C_{p}. \label{eq3}
\end{equation}
Eq. (3) qualitatively agrees with the experimental result shown in
figure 2(d). At low frequencies, the first term is important. Thus
tan$\delta$ decreases with $f$. At high frequencies, however, the
second term becomes predominant. As a result, tan$\delta$
increases almost linearly with the frequency for all the samples
when $f \geq 10^{5}$.

Eq. (3) clearly shows that tan$\delta$ can be lowered by either
increasing $R_{ins}$ or decreasing $R_{sc}$. Then, which factor is
responsible for the decrease of tan$\delta$ through the La-for-Ca
substitution?

Fortunately, the $R_{sc}$ and $R_{ins}$ values can be obtained by
an impedance spectrum analysis.\cite{Sinclair} As shown in the top
right inset of figure 3, the non-zero intercept on the $Z'$ at the
high frequency give the $R_{sc}$ value. On the other hand,
$R_{ins}$ might also be obtained by another intercept on the $Z'$
axis at very low frequency. Since the experimental impedance data
at 270 K only cover a part of the arcs for the limit of the
measured frequency range, $R_{ins}$ was then determined by a data
fitting and extrapolation using the well-known Cole-Cole empirical
relation.

\begin{figure}[tbp]
\includegraphics[width=7.5cm]{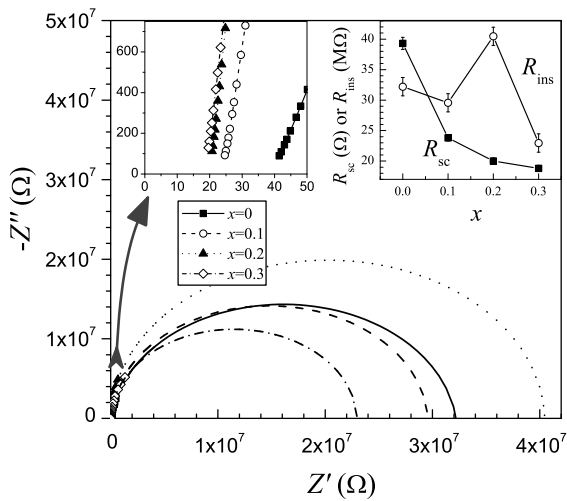}
\caption{$Z^{'} - Z^{''}$ plot on the impedance complex plane at
270 K for the Ca$_{1-x}$La$_{x}$Cu$_{3}$Ti$_{4}$O$_{12}$ ceramics.
The lines are the fitted curves using the Cole-Cole empirical
relation. The top right inset shows an enlarged view for the high
frequency data close to the origin. The top left inset plots the
resistance of the semiconducting region ($R_{sc}$) and that of the
insulating barriers ($R_{ins}$) as a function of La content.}
\end{figure}

As can be seen in the top left inset of figure 3, $R_{sc}$ does
decease with the La substitution. Furthermore, the decrease is
very sharp at the low $x$ value, and it becomes mild when $x \geq$
0.2. This is in conformity with the decrease of tan$\delta$ in
figure 2(b) (Note that the second term of tan$\delta$ in Eq. (3)
predominates at 10 kHz). On the other hand, $R_{ins}$ does not
changes so much with increasing $x$. It is also noted that the
decrease of $C_p$ (or $\varepsilon_r$) cannot explain the large
magnitude of the drop in tan$\delta$. Therefore, we conclude that
the notable decrease of tan$\delta$ by the La substitution is
mainly due to the decrease in resistivity of the CCTO
grains/subgrains. The La-for-Ca substitution is expected to induce
excess electrons due to the charge compensation effect, which
basically explains the decrease of $R_{sc}$.

One may note that tan$\delta$ at low frequencies does not decrease
so much (figure 2(d)). The sample of $x$ = 0.3 shows even higher
tan$\delta$ than that of the parent compound at 20 Hz, primarily
due to the lower $R_{ins}$ value. To decrease the low-frequency
tan$\delta$, the direct route is to increase the resistivity of
the barriers. Very recently, we succeeded in increasing $R_{ins}$
(and thus decreasing tan$\delta$) by incorporating additional
CaTiO$_3$.\cite{Yan} The decrease of tan$\delta$ was also reported
in the CaCu$_{3}$Ti$_{4}$O$_{12}$/CaTiO$_3$ diphasic
samples.\cite{Kobayashi} It is expected that the combination of
the two routes may further optimize the dielectric property in the
CCTO material.

In summary, we have developed a novel route to improve the
dielectric properties of the CCTO material based on the IBLC
model. By the La-for-Ca substitution in CCTO the loss tangent was
suppressed remarkably while the giant dielectric constant still
remains. The impedance spectroscopy analysis shows that the
conductivity of the semiconducting regions increases substantially
with the La substitution while the internal barriers remain highly
resistive, leading to the notable decrease in the loss tangent. We
hope that the present work will be able to promote the
CCTO-related materials to practical applications.

This work was supported by the National Science Foundation of
China (Grant No. 10104012).

\end{document}